# Molecular dynamics studies on the NMR structures of rabbit prion protein wild-type and mutants: surface electrostatic charge distributions

(Version 1)

**Jiapu Zhang**[ab*], **Feng Wang**[a]

[a]Molecular Model Discovery Laboratory, Department of Chemistry & Biotechnology, Faculty of Science, Engineering & Technology, Swinburne University of Technology, Hawthorn, Victoria 3122, Australia;

[b]Graduate School of Sciences, Information Technology and Engineering & Centre of Informatics and Applied Optimisation, Faculty of Science, The Federation University Australia, Mount Helen Campus, Mount Helen, Ballarat, Victoria 3353, Australia

[*]Correspondence address: Tel: +61-3-9214 5596, +61-3-5327 6335, +61-423 487 360; E-mails: jiapuzhang@swin.edu.au, j.zhang@federation.edu.au, jiapu_zhang@hotmail.com

**Abstract:**

Prion is a misfolded protein found in mammals that causes infectious diseases of the nervous system in humans and animals. Prion diseases are invariably fatal and highly infectious neurodegenerative diseases that affect a wide variety of mammalian species such as sheep and goats, cattle, deer, elk and humans etc. Recent studies have shown that rabbits have a low susceptibility to be infected by prion diseases with respect to other animals including humans. The present study employs molecular dynamics (MD) means to unravel the mechanism of rabbit prion proteins (RaPrP$^C$) based on the recently available rabbit NMR structures (of the wild-type and its two mutants of two surface residues). The electrostatic charge distributions on the protein surface are the focus when analysing the MD trajectories. It is found that can conclude that surface electrostatic charge distributions indeed contribute to the structural stability of wild-type RaPrP$^C$; this may be useful for the medicinal treatment of prion diseases.

**Abbreviations**: PrP$^C$, a soluble normal cellular prion protein; PrP$^{Sc}$, insoluble abnormally folded infectious prions; BSE, bovine spongiform encephalopathy; TSE, Transmissible

Spongiform Encephalopathy; CJD, Creutzfeldt-Jakob Disease; vCJD, variant Creutzfeldt-Jakob Disease; saPMCA, *s*erial *a*utomated Protein Misfolding Cyclic Amplification.

## Introduction

It has been a challenge to rational whether the contagious Transmissible Spongiform Encephalopathy (TSE) is caused by prions (Prusiner, 1997 & 1998; Soto & Castilla, 2004; Soto, 2011; Fernandez-Borges et al., 2012). As a misfolded protein, prion is neither a virus, nor a bacterium, and nor any microorganism. Prion disease cannot be caused by the vigilance of the organism immune system but it can freely spread from one species to another species. Humans TSEs (for example, Creutzfeldt-Jakob Disease (CJD) and variant CJD (vCJD)) can happen randomly through a number of processes, such as infections of transplanted tissue, blood transfusions and/or consumption of infected beef products, etc. Many mammals such as cat, mink, deer, elk, moose, sheep, goat, nyala, oryx, greater kudu and ostrich etc. are also susceptible to TSEs. However, a small group of other animals such as rabbits, horses and dogs seem to be little affected by prions (Vorberg, Martin, Eberhard, & Suzette, 2003; Khan et al., 2010; Polymenidou et al., 2008; Zhang, 2011a; Zhang & Liu, 2011). As a result, it is important to understand and to identify the specific causes why these animals are unlikely to be affected by prions, as it will provide insight to prion diseases and help to resolve the prion diseases issue.

The role of $PrP^{Sc}$ infection in animals such as rabbit has been subject to a heated debating. Although a number of previous studies showed that a few animals such as rabbits exhibit low susceptibility to be infected by the $PrP^{Sc}$ (Vorberg, Martin, Eberhard, & Suzette, 2003; Khan et al., 2010; Barlow & Rennie, 1976; Fernandez-Funez et al., 2009; Korth et al., 1997; Courageot et al., 2008; Vilette et al., 2001; Nisbet et al., 2010; Wen et al., 2010a; Wen et al., 2010b; Zhou et al., 2011; Ma et al., 2012), other recent studies do not agree. Rather, recent studies suggested that (i) the rabbit prion may be produced through saPMCA (serial automated Protein Misfolding Cyclic Amplification) in vitro (Chianini et al., 2012; Fernandez-Borges et al., 2012), and (ii) the rabbit prion is infectious and transmissible (Chianini et al., 2012). In their most recently studies, Zhang et al (2014) and Vidal et al. (2013) suggested that rabbits were not challenged directly in vivo by other known prion strains. In addition, the saPMCA results did not pass the test of the known BSE strain of cattle. As a result, rabbits can be considered as a prion resistant species.

A limited number of experimental structural data for RaPrP$^C$ are available from the Protein Data Bank (PBD: http://www.rcsb.org/). For example, the structures of RaPrP$^C$ obtained from NMR (Li et al., 2007, with a PBD entry of 2FJ3) and X-ray (Khan et al., 2010, with a PBD entry of 3O79) measurements. As a result, it is desirable to reveal the properties and specific mechanisms of the rabbit PrP$^C$ (RaPrP$^C$) and the conversion process of PrP$^C$→PrP$^{Sc}$ of rabbit from limited experimental results. Here PrP$^C$ is a soluble normal cellular prion protein and PrP$^{Sc}$ is insoluble abnormally folded infectious and diseased prions. The present study will base on the NMR structure of RaPrP$^C$ (and its I214V and S173N mutants: NMR structure PDB entries 2JOM, 2JOH, respectively) using molecular dynamics (MD) simulation techniques. The information from the present MD studies is able to provide valuable insight for the PrP$^C$→PrP$^{Sc}$ conversion. The information will provide useful rational in the design of novel therapeutic approaches and drugs that stop the conversion and disease propagation.

This paper can capture the clear α-helices→β-sheets conversion of PrP$^C$→PrP$^{Sc}$ under pH environments from neutral to low. Salt bridge (SB) changes of charged residues under these two environments can lead to this conversion. Thus, electrostatic charges of charged residues are focused to study in this paper.

In this study, electrostatic potential (ESP) surfaces of the proteins are calculated in order to determine the shape of the proteins, which helps to reveal how the ligands (drugs) dock into the proteins, the lock and key model in computer-aided drug design (Zhang, 2012). That is, the ESP surfaces are used as indicators for different inhibitors with substrates or transition states of the reaction. These bioinformatics results may be useful to understand the mechanisms and to suggest helpful information in the medicinal treatment of prion diseases. As a result, the analysis of surface electrostatic charge distributions of MD trajectories will be presented. The rest of this paper is organized as follows. The MD simulation materials and methods for NMR structures of RaPrP$^C$ wild-type and mutants are provided in next section, followed by the analysis and discussion focusing on the MD trajectory results of the surface electrostatic charge distributions and their discussions. Finally, concluding remarks on RaPrP are summarized.

## Materials and methods

The materials, e.g., data used in the present study are based on the NMR PDB files of 2FJ3.pdb, 2JOM.pdb, 2JOH.pdb. The MD methods employed are the same as the previous studies (Zhang et al., 2014; Zhang, 2010). Briefly, all simulations used the ff03 force field of the AMBER 11 package (Case et al., 2010). The systems were surrounded with a 12 Å layer of TIP3PBOX water molecules and neutralized by sodium ions using the XLEaP module of AMBER 9. Then 14 Cl-, 14 Cl-, 13 Cl- and 5909, 4185 and 4729 water molecules were added in the rabbit prion wild-type, S173N mutant and I214V mutant, respectively for pH < 7. To remove the unwanted hydrogen bond contacts, the systems of the solvated proteins with their counter ions were minimized mainly by the steepest descent method and followed by a small number of conjugate gradient steps on the data. Next, the solvated proteins were heated from 100 K to 450 K step by step in a 3 ns duration (prions are not completely removed even at a high temperature of 600 ℃ (Rhee et al., 2009). For prions many experimental works and molecular dynamics works have done at 500 K (e.g. Sekijima et al., 2003); simulation results on protein structures and their dynamics of this paper showed that the force field parameters are suitable to allow simulations at 450K). Three sets of initial velocities denoted as seed1, seed2 and seed3 are performed in parallel for stability. The thermostat algorithm used is the Langevin thermostat algorithm in constant NVT ensembles. The SHAKE algorithm (only on bonds involving hydrogen) and PMEMD (Particle Mesh Ewald Molecular Dynamics) algorithm with non-bonded cutoff of 12 Å were used during heating. Equilibrations were reached in constant NPT ensembles under Langevin thermostat for 5 ns. After equilibrations, production MD phase was carried out at 450 K for 30 ns using constant pressure and temperature ensemble and the PMEMD algorithm with the same non-bonded cutoff of 12 Å during simulations. The step size for equilibration was 0.5 fs and 1 fs in the MD production runs. The structures were saved to file every 1000 steps. The MD simulations are repeated under neutral pH environment.

In order to obtain the low pH (acidic) environment, the residues HIS, ASP, GLU were changed into their zwitterion forms of HIP, ASH, GLH, respectively, and Cl- ions were added by the XLEaP module of the AMBER package. Thus, the SBs of the system (residues HIS, ASP, GLU) under the neutral pH environment were broken in the low pH environment (zwitterion forms of HIP, ASH, GLH). At acidic pH, our MD simulations can also capture the misfolding of prion proteins from RaPrP$^C$ α-helices to RaPrP$^{Sc}$ β-sheets (Cheng et al., 2014).

**Discussion on the MD methods**

The pmemd.MPI is faster and scales better in parallel than sander.MPI, but pmemd is less accurate than sander. The main reason for choosing PMEMD (Salomon-Ferrer et al., 2013) in our case is that (i) the solvated systems are large, (ii) the MD simulation time are long, and (iii) the parallel supercomputer resources of VLSCI (http://www.vlsci.org.au).

The use of cutoffs in GB (Generalized Born) simulations as implemented in PMEMD does not conserve energy, and their use involves an approximation with an unknown effect on accuracy. For this reason, we chose not to implement van der Waals (vdW) and electrostatic cutoffs in the GPU version of this code. Cutoffs for the nonbonded interactions are implemented for explicit solvent simulations with periodic boundary conditions using the PME method, as described in (Götz et al., 2012). However, cutoffs in calculating the effective Born radii are supported.

We explain a bit more about PME and the explicit water with counter ions distribution as described in (Walker R., 2010). In the periodic boundary method the particles being simulated are enclosed in a box which is then replicated in all three dimensions to give a periodic array. During the simulation only one of the particles is represented, but the effects are reproduced over all the image particles with each particle not only interacting with the other particles but also with their images in neighbouring boxes. Particles that leave one side of the box re-enter from the opposite side as their image. In this way the total number of particles in the central box remains constant. The periodic boundary method would appear to be very computationally intensive. However, by employing the PME method it is possible to obtain the infinite electrostatics in a way that scales as NlnN time. This involves dividing the calculation up between a real space component and a reciprocal space component. The use of PME means we are calculating all of the 'infinite' electrostatics and so are not actually truncating the electrostatics. The VDW interactions are still needed though which means we cannot make the cut-off too small less than 8 Å for periodic boundary PME calculations. Prior to MD, we use optimization (first Holding the solute fixed to just minimize the positions of the water and ions and then minimizing the entire system) to relax the system to remove bad VDW and electrostatic interactions and the gaps between the solvent and solute and solvent and box edges, etc. During the equilibration period, we first run short MD at constant volume and then use periodic boundaries to keep the pressure constant and so allow the volume of the box to change. This will allow the water and ions to equilibrate around the solute and come to an equilibrium density. The visual description of colours for charges should be averaged from residue charge.

We have chosen 450 K, which is high temperature to SB. This means that the Cp is low enough to keep the initial structure (ΔG<0) at 450K (Chan et al., 2011); but it is not simulated using the Gibbs-Helmholtz equation (Chan et al., 2011) by all versions of the Amber package (http://ambermd.org/).

We use FirstGlance in Jmol (http://bioinformatics.org/firstglance/fgij/) to detect all the charges of RaPrPC wild-type, I214V mutant and S173N mutant: 10+ (7 ARG, 3 LYS) (3HIS) and 13- (5 ASP, 7 GLU, 1 C-terminus), in defining the following SBs that we have calculated: ASP146-ARG147, GLU210-ARG207, GLU206-LYS203, GLU206-ARG207, ASP146-HIS139, GLU151-ARG150, GLU151-ARG147, GLU151-ARG155, ASP177-ARG163, GLU145-ARG135, ASP143-HIS139, GLU145-HIS139, GLU195-ARG155, ASP146-ARG150, ASP177-HIS176, ASP143-ARG147, GLU195-LYS193, HIS186-ARG155, HIS186-LYS184, ASP166-ARG163, GLU210-HIS176, HIS139-ARG135, GLU199-LYS193, GLU206-HIS176 (Zhang et al., 2014).

This paper is focusing to study the ESP surface charge distributions. Here in the below we will specially discuss some excellent dielectrique constant (Chan et al., 2011) and dielectric constant (Guest et al., 2011) for the calculations of ESP surface charges.

**Discussion on dielectrique constant and free energy**

The dielectric response for a particular region of the protein is obtained at the microscopic level by employing a TI/MD (Thermodynamic Integration / Molecular

Dynamics) method; MD simulations and "TI" are used to determine how the protein responds to charge insertion at a particular site at the atomistic level. In this paper we presented the external polarisation in terms of charges (positives, neutral, negatives). This paper uses Maestro 9.7 2014-1 (Academic use only) free package (http://www.schrodinger.com/) to draw the Poisson-Boltzmann ESP surface charges of the external polarisation: we choose indi=1.0 as the solute dielectric constant, exdi=80.0 as the solvent dielectric constant, and 12 Å as the solvent radius, 450 K as the Temperature, and EPS mapped on molecular surface is chosen. We know that the delta $\Delta Cp \Rightarrow \Delta Gu=f(T)$ is a good method to show how a molecular structure is more stable with a SB (Figure S4 of (Chan et al., 2011)); using the delta method on the 2FJ3.pdb file of the rabbit PrP, the distance of ASP177-ARG163 (O−N) gives a SB of 10 Å which is nearly null in terms of energy. But this is just the result of one snapshot under special conditions; during the whole MD simulations at temperatures such as 300 K, 350 K, 400 K, 450 K, and 500 K the distance of ASP177-ARG163 (O−N) is at the SB distance during most of the 30 ns' MD simulation time.

**About dielectric constant**

In the method using a mesoscopique dielectric constant a nice figure (Figure 4B of (Guest et al., 2011)) shows the change of ε which gives what is under the coloured surface. The transparent surfaces shows contours of equal dielectric as determined from heterogeneous mesoscopique dielectric theory (Guest et al., 2011). SBs into the core of the protein will show a better energy for stabilization (weak ε, therefore better $\Delta G_u$) than all SBs rather on the external protein. All the charges we are showing as red and blue are not just on the periphery. SBs are calculated by oppositely-charged atoms that are within 6.5 Å and are not directly hydrogen-bonded. It should be the average charge calculated per residue or the specific atom charge of the residue. The donor residues involved are ASP, GLU, HID and the acceptor residues involved are LYS, ARG, HID, and the real computed distance is within 6.5 Å in Amber package. On prion proteins, Guest et al. (2011) declared "Salt bridges known to be absent in disease causing human mutants of the prion protein were found to be among the strongest present in the protein, so that the human mutants were electrostatically the least stable of those proteins studied. Conversely, the prion protein with the most stable salt bridges belonged to a species known to be resistant to prion disease (frog)" (Guest et al., 2010 & 2011).

# Results and discussion

The MD results from the simulations are summarized in Figures 1, 2 and 3, respectively, under the parallel conditions of seed1, seed2 and seed3 in the duration of 30 ns. Each figure, say, Figure 1, it composes of six panels --- three on the left for the system under the neutral pH condition and three on the right for the systems under the acidic (low pH) solutions. The variations of the secondary molecular structures of the wild-type (RaPrP$^C$) and its I214V and S173N mutants are given on the top row, middle row and bottom row, respectively. As a result, the six panels in Figure 1 can be called as 1N1, 1N2, 1N3, 1L1, 1L2 and 1L3, for the results under seed1 conditions, in neutral pH environment for RaPrP$^C$, I214V mutant and S173 mutant for 1N1, 1N2 and 1N3; in low pH environment (acidic) for RaPrP$^C$, I214V mutant and S173 mutant for 1L1, 1L2 and 1L3. For example, 1L2 is the results for seed1 conditions in low pH solution for I214V mutant. Within each panel, say, 1N1 (the top-left corner panel), three subpanels named H1, H2 and H3 from bottom to top represents the α-helix 1, α-helix 2 and α-helix 3 of the prion protein, respectively.

Under the convention, coloured codes represent various helices. The information in the figures was produced using the DSSP program (Kabsch & Sander, 1983). For example, red, green, yellow, blue, dark blue, purple and orange represent for H, I, G, B, E, T and S, accordingly. The coloured codes represent various helices. For example, letter H refers to the α-helix, I refers to the π-helix, G refers to the 3-helix or 3/10 helix, B refers to the residue in isolated β-bridge, E refers the extended strand (participates in β-ladder), T refers to the hydrogen bonded turn, and S refers to the bend. The colour coded letters are given in the figures.

As can be seen from Figures 1-3, the top-left corner panels, i.e, 1N1, 2N1 and 3N1 are dominated by red colour, indicating that under the neutral pH conditions, the three α-helices (H1, H2 and H3) of the wild-type prion RaPrP$^C$, remain dominant the α-helices without significant changes during the period of 30 ns, regardless the seed conditions. For example, the α-helices (H1, H2 and H3 in red colour) of 1N1 in Figure 1 (top-left panel) does not experience any apparent colour changes, indicating that under the neutral pH condition, the α-helices of the wild-type prion resist structural changes. This is particular the case in H3 and H2, although small noticeable changes in H1 has been observed. However, the I214V (second rows) and S173N (third rows) mutants of the wild-type prion protein show significant changes indicated by their changes of colour codes. This is also the cases under acidic condition (i.e., low pH environment) on the right hand side of the figures, which shows colourful panels---indication very little similarities can be seen in the corresponding bands. As Figures 1, 2 and 3 are under parallel conditions with similar changes patterns, the following discussion will focus on Figure 1 under seed1 conditions.

In Figure 1, the 1N1 panel is unique---it is dominant by red, indicating little mutational changes, whereas all other panels in this figure results in colourful changes. It reveals that the three α-helices of the wild-type (RaPrP$^C$) prion protein are unfolded and remain in the same structures under neutral pH conditions. However, in low pH environment, this wild-type prion protein (1L1) turns into a colourful panel of almost all of the seven colours but large presentation of orange (hydrogen bonded bend). Therefore, under acidic condition (low pH environment), the SB network of the wild-type (RaPrP$^C$) is broken thus leads to the unfolding of the stable α-helical structures of RaPrP$^C$ (Zhang et al, 2014). Hence, it suggests that the structural distributions of the wild-type (RaPrP$^C$) protein depends on the pH. On the other hand, the I214V and S173N mutants under either neutral or low pH environment, such as 1N2/1L2 and 1N3/1L3, are unfolded into rich β-sheet structures. Generally, we could see the clear secondary structure changes from (α-, π-, 3$_{10}$-) helix structures to β-sheet (ladder and bridge) structures from Figures 1~3.

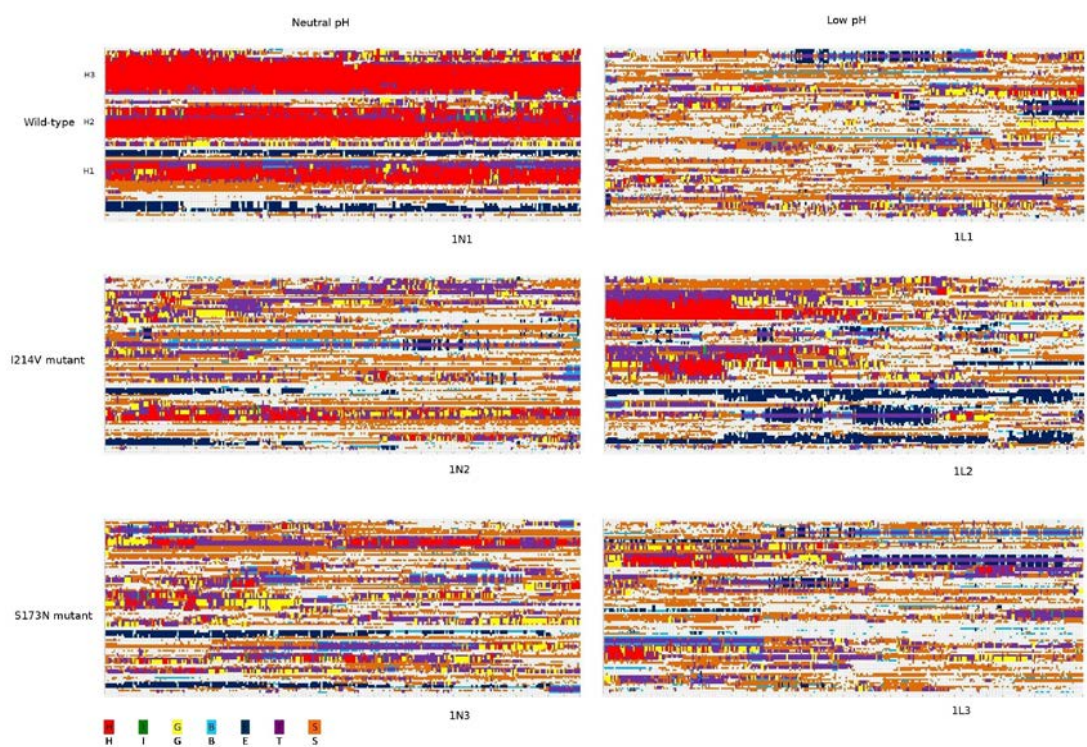

**Figure 1**: Seed1. Secondary Structure graphs for RaPrP$^C$ wild-type, I214V mutant and S173N mutant (from up to down) at 450 K (x-axis: time (0–30 ns), y-axis: residue number (124–228); left column: neutral pH, right column: low pH. H is the α-helix, I is the π-helix, G is the 3-helix or 3/10 helix, B is the residue in isolated β-bridge, E is the extended strand (participates in β-ladder), T is the hydrogen bonded turn, and S is the bend.

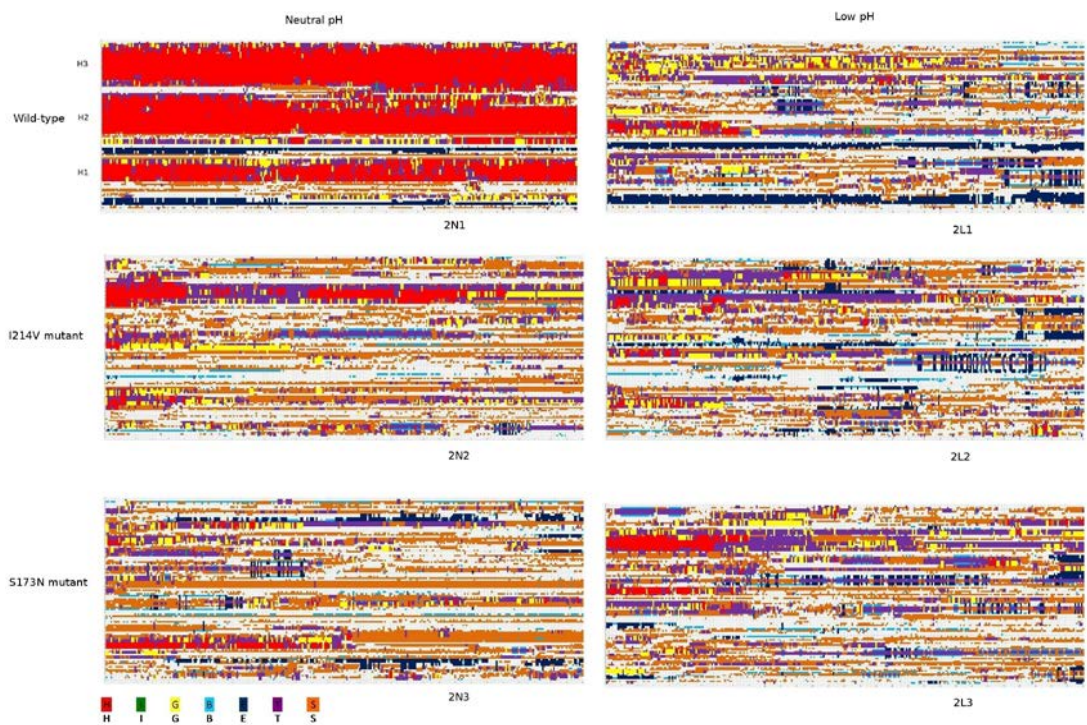

**Figure 2**: Seed2. Secondary Structure graphs for RaPrP$^C$ wild-type, I214V mutant and S173N mutant (from up to down) at 450 K (x-axis: time (0–30 ns), y-axis: residue number (124–228); left column: neutral pH, right column: low pH. H is the α-helix, I is the π-helix, G is the 3-helix or 3/10 helix, B is the residue in isolated β-bridge, E is the extended strand (participates in β-ladder), T is the hydrogen bonded turn, and S is the bend.

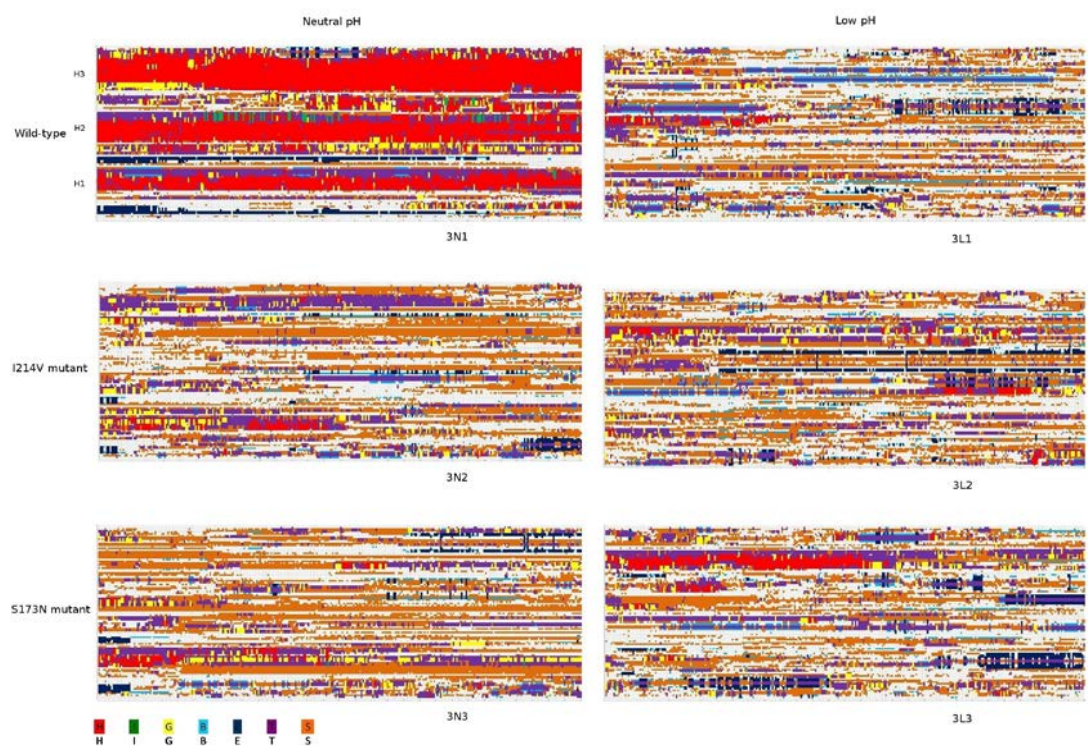

**Figure 3**: Seed3. Secondary Structure graphs for RaPrP$^C$ wild-type, I214V mutant and S173N mutant (from up to down) at 450 K (x-axis: time (0–30 ns), y-axis: residue number (124–228); left column: neutral pH, right column: low pH. H is the α-helix, I is the π-helix, G is the 3-helix or 3/10 helix, B is the residue in isolated β-bridge, E is the extended strand (participates in β-ladder), T is the hydrogen bonded turn, and S is the bend.

Residue at 214 within the C terminus could potentially influence the conversion of RaPrP$^C$ to RaPrP$^{Sc}$ (Vorberg et al., 2003). For residue at 173 in the β2-α2 loop, Wen et al. (2010) reported a large-continuous-positive-charge-surface ESP, which is unique to the RaPrP$^C$ prion protein (Wen et al, 2010b). In the studies, the solution structures (backbones) of the RaPrP$^C$ protein and its S173N variant are determined for their structured C-terminal domains. It suggested that the highly ordered β2-α2 loop has been well recognized by the NMR-signals contribute to the structural stability of RaPrP$^C$ (Wen et al., 2010a,b) at one snapshot of 300 K and 1 ATM condition. In Figure 4, the surface EPS charge distributions are calculated based on the NMR structure of the RaPrP$^C$ wild type.

Figures 5 and 6 show the surface ESP charge distributions of the wild-type RaPrP$^C$ protein at a temperature of 450 K for the neutral pH environment obtained in the present study. Figure 5 gives the snapshots at 5 ns, 10 ns, 15 ns, 20 ns, 25 ns, 30 ns, for the conditions of seed1 (first row), seed2 (second row) and seed3 (3$^{rd}$ row). In the figure, the charges coloured by red and blue represent negative and positive respectively. As shown in Figure 5, it is discovered that a large region of positive charge (blue) distribution has still been reserved during the long MD simulation of 30 nanoseconds. However, the continuous of the land has not been always reserved yet (where continuous means connect to each other, not discreted), as indicated by Figure 6 with rotational views.

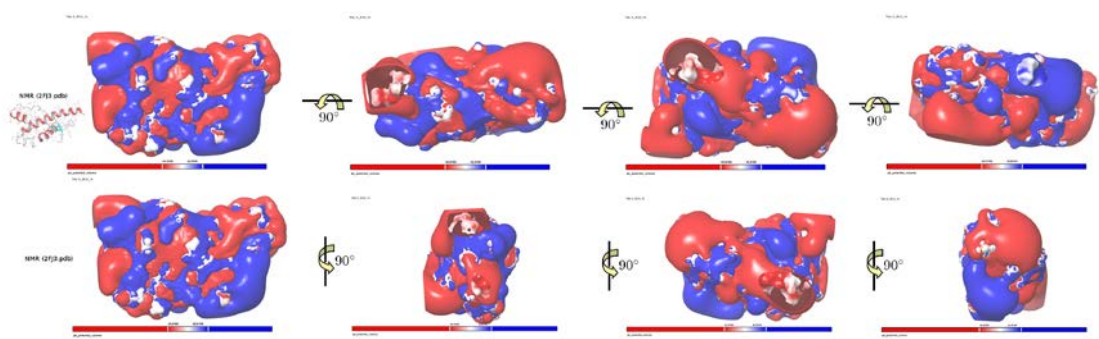

**Figure 4:** Surface electrostatic charge distributions for the NMR structure of RaPrP$^C$ wild-type, where blue is for positive charge whereas red is for negative charge. The pb_potential_volume is ±43.8195.

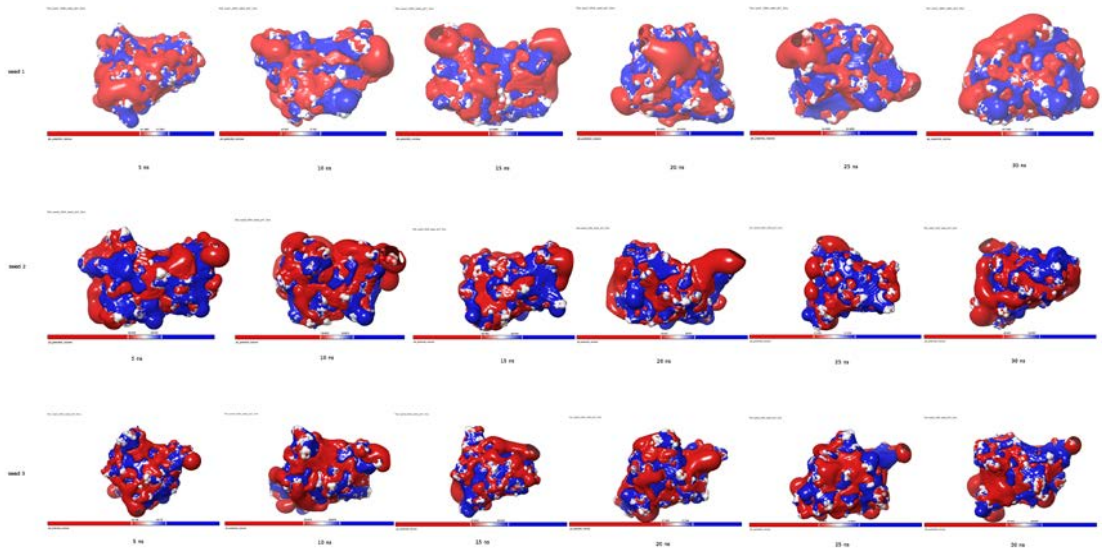

**Figure 5:** Surface electrostatic charge distributions for RaPrP$^C$ wild-type at 450 K in neutral pH environment at 5 ns, 10 ns, 15 ns, 20 ns, 25 ns, 30 ns (left to right), where blue is for positive charge whereas red is for negative charge. Up to down: seed1~seed3. The pb_potential_volumes for seed1 are ±21.3561 (5 ns), ±27.637 (10 ns), ±23.9485 (15 ns), ±28.2425 (20 ns), ±32.4692 (25 ns), ±25.1583 (30 ns), for seed2 are ±26.452 (5 ns), ±19.6673 (10 ns), ±28.1642 (15 ns), ±25.912 (20 ns), ±31.4185 (25 ns), ±29.0557 (30 ns), and for seed3 are ±24.138 (5 ns), ±28.8575 (10 ns), ±26.5422 (15 ns), ±21.7444 (20 ns), ±31.0833 (25 ns), ±28.0295 (30 ns).

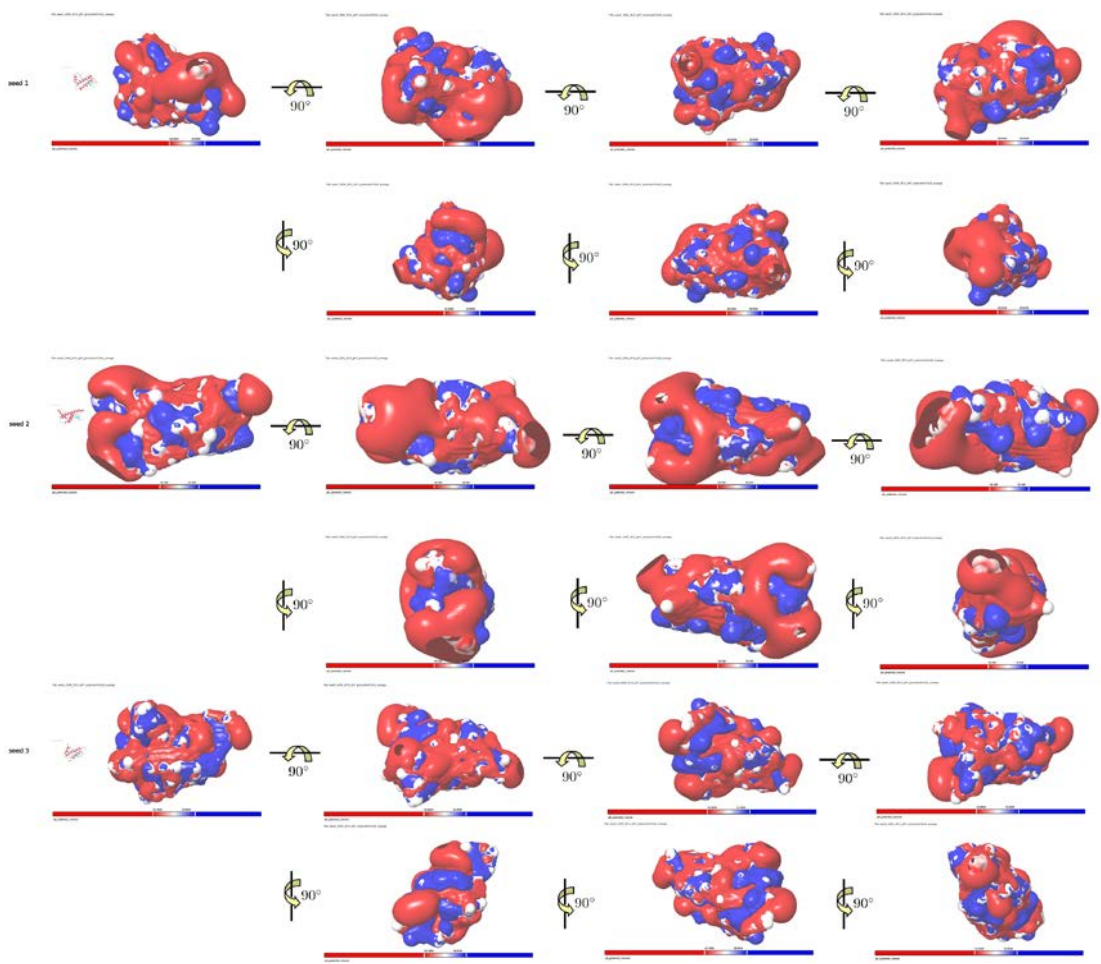

**Figure 6:** Surface electrostatic charge distributions for the average structure of 30 ns' MD of RaPrP$^C$ wide-type at 450 K in neutral pH environment, where blue is for positive charge whereas red is for negative charge. Up to down: seed1~seed3. The pb_potential_volumes for seed1, seed2, seed3 are ±25.8555, ±25.328, ±33.5938 respectively.

The I214V mutation does not change the ESP due to the non-charged feature of the two amino acids at site 214 (Wen et al., 2010a). Figure 7 shows the surface ESP charge distributions of the mutation of the wild-type RaPrP$^C$ protein at a temperature of 450 K in the neutral pH environment obtained in the present study. However, it reveals that the structures of I214V mutant do not unfold from the α-helices (refer to Figures 1~3) in the 30 ns duration of the MD simulation. This suggests that the ESP may be not the reason leading to the structure unfolding of the I214V mutant of the RaPrP$^C$ protein.

The S173 mutation, however, just lies in the β2-α2 loop. Thus this mutation S173N leads to the changes of the network of the ESP in this loop, which is indicated in Figure 8 under seed1 and seed2 conditions, which is very different from seed3 conditions in the same figure. It indicates that the ESP may lead to the structure unfolding of the S173N mutant of the RaPrP$^C$ protein in seed1 and seed2, whereas the ESP may be not the reason for the structure unfolding of the S173N mutant of the RaPrP$^C$ protein. From Figures 6~8, we may see that the pb_potential_volume for wild-type is less than for mutants.

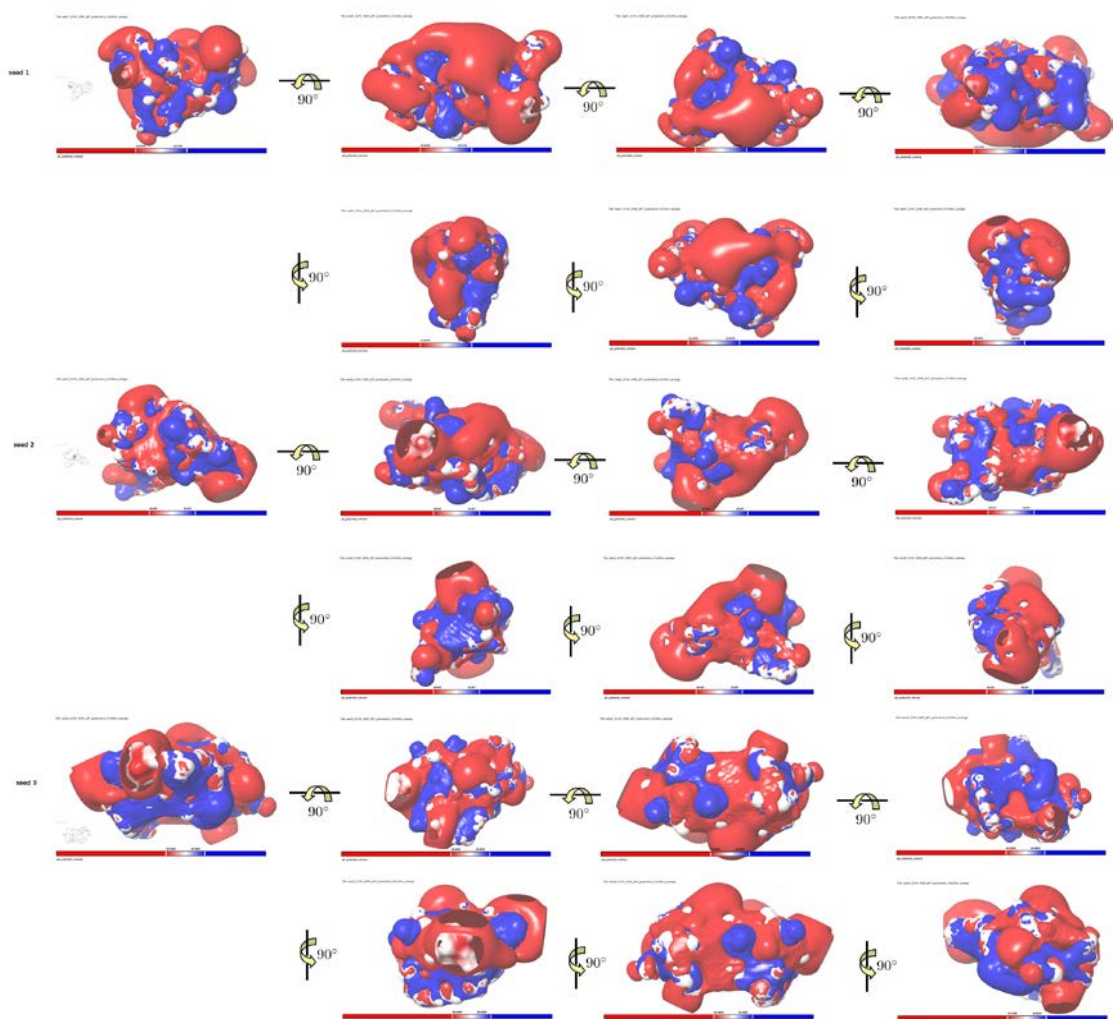

**Figure 7:** Surface electrostatic charge distributions for the average structure of 30 ns' MD of RaPrP$^C$ I214V mutant at 450 K in neutral pH environment, where blue is for positive charge whereas red is for negative charge.

Up to down: seed1~seed3. The pb_potential_volumes for seed1, seed2, seed3 are ±34.9142, ±38.007, ±40.6668 respectively.

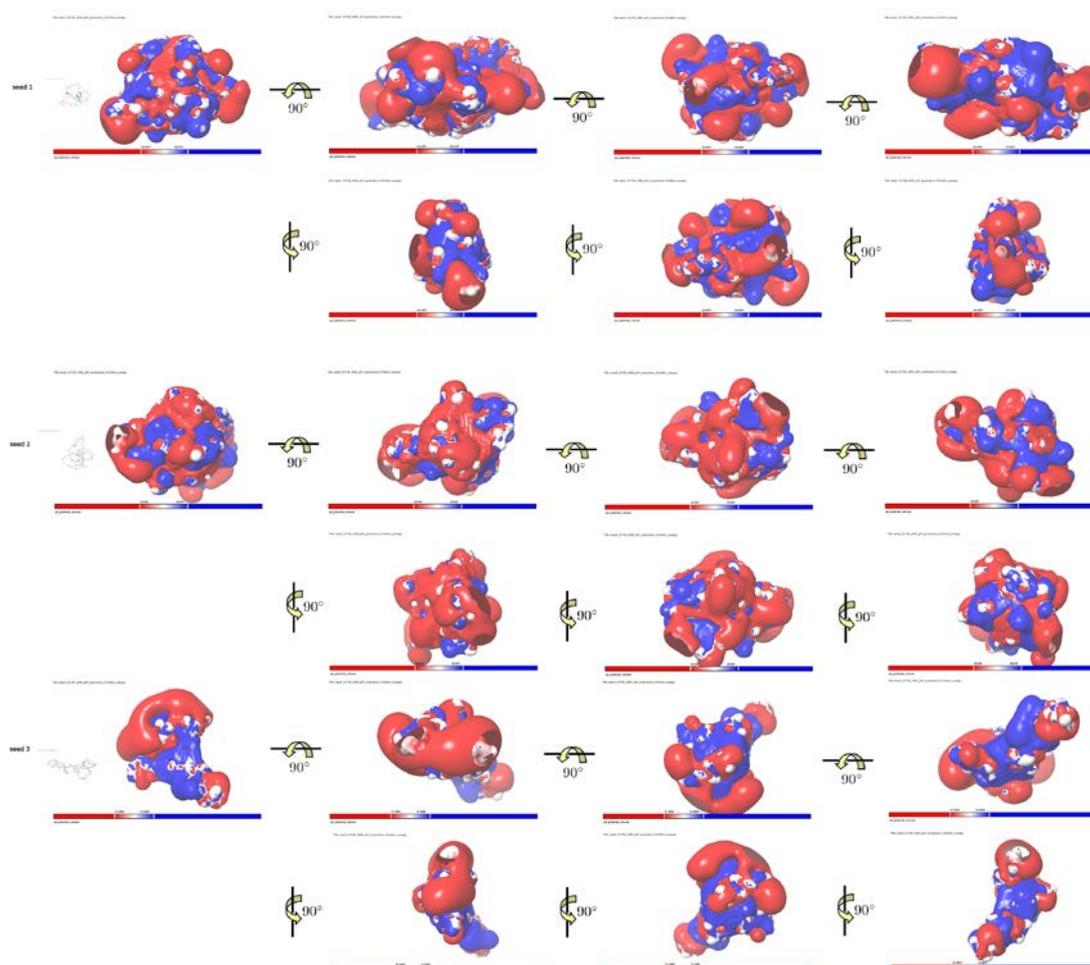

**Figure 8:** Surface electrostatic charge distributions for the average structure of 30 ns' MD of RaPrP$^C$ S173N mutant at 450 K in neutral pH environment, where blue is for positive charge whereas red is for negative charge. Up to down: seed1~seed3. The pb_potential_volumes for seed1, seed2, seed3 are ±42.8397, ±39.629, ±31.4955 respectively.

# Conclusions

There are still a number of challenges remain in "prion" theory. The present study using the molecular dynamic simulation strategies has found that surface electrostatic charge distributions play an important role in the structural stability of RaPrP$^C$ but are clearly not the only reason contributing to the structural stability the RaPrP$^C$ protein. This finding only partially agrees to the conclusion that "large-continuous-positive-charge-surface ESP contributes to the structural stability of RaPrP$^C$" (especially in the β2-α2 loop region) (Wen et al., 2010a,b). Further studies to understand the specific mechanism of RaPrP$^C$ are still needed.

# Acknowledgments

One of the authors (JPZ) thanks Dr. Marawan Ahmed for his suggestion to use the Maestro 9.7 (Academic Use Only) to draw the surface electrostatic charge graphs. This research has been supported by a Victorian Life Sciences Computation Initiative (VLSCI) grant numbered VR0063 on its Peak Computing Facility at the University of Melbourne, an initiative of the Victorian Government of Australia. Swinburne University G2 supercomputing facilities are also acknowledged.